\documentclass[aps,prc,twocolumn,amsmath,superscriptaddress,floatfix,nofootinbib]{revtex4-2}

\usepackage{amsmath,amssymb,amsfonts,mathtools,bm}
\usepackage{graphicx}
\usepackage{xcolor,ulem}
\usepackage{braket}
\usepackage{ragged2e} 
\usepackage{changes}

\usepackage[caption=false]{subfig}
\usepackage{hyperref}
\hypersetup{
bookmarksnumbered=true,bookmarksopen=true,colorlinks=true,    citecolor=blue,linkcolor=red,urlcolor=blue
}
\usepackage{subfig} 

\usepackage{graphicx}

\graphicspath{
    {fig/}
}

\begin{document}

\title{Triangular and fourth-order flow of D mesons in Pb-Pb collisions with event-by-event fluctuations
}

\author{Yiyun Tang}
\affiliation{Department of Physics, Tianjin University, Tianjin 300354, China}
\author{Jiamin Liu}
\affiliation{Department of Physics, Tianjin University, Tianjin 300354, China}

\author{Yunpeng Liu}
\affiliation{Department of Physics, Tianjin University, Tianjin 300354, China}

\author{Baoyi Chen}\email{baoyi.chen@tju.edu.cn}
\affiliation{Department of Physics, Tianjin University, Tianjin 300354, China}
\affiliation{International Joint Institute of Tianjin University, Fuzhou,  Tianjin University, Tianjin  300072, China}

\date{\today}

\begin{abstract} 

Charm quarks are valuable probes of the transport properties of the quark-gluon
plasma (QGP), yet the role of event-by-event medium fluctuations in shaping
their observables remains to be quantified. We study this problem in Pb-Pb
collisions at $\sqrt{s_{NN}} = 5.02$~TeV by coupling the iEBE-VISHNU
event-by-event viscous hydrodynamic framework to a Langevin transport model
with instantaneous coalescence hadronization. We find that fluctuations have
little effect on the azimuthally averaged $D$ meson nuclear modification
factor $R_{AA}$, which remains close to the smooth-background baseline with
only modest momentum-dependent deviations. In contrast, the anisotropic flow
coefficients $v_n$ are increasingly sensitive to fluctuations with rising
harmonic order $n$. In particular, the triangular flow $v_3$, which vanishes
in the smooth background, is entirely generated by initial-state fluctuations,
and our results reproduce the main features of the available CMS measurements.
We further present predictions for the $D$ meson $v_4$, which provide a
quantitative benchmark for future measurements of higher-order heavy-flavor
flow and help elucidate the interplay between event-by-event QGP fluctuations
and charm quark dynamics.

\end{abstract}

\maketitle

\section{Introduction}
Collisions of heavy nuclei at ultrarelativistic energies create strongly
interacting matter under conditions of extreme temperature and energy density,
providing access to the deconfined regime of quantum chromodynamics
(QCD)~\cite{Harris:1996zx,Braun-Munzinger:2007edi}. At the small baryon
chemical potentials realized in these collisions, lattice QCD establishes the
transition from hadronic matter to the quark-gluon plasma (QGP) as a smooth
crossover~\cite{Rischke:2003mt,Aoki:2006we}. Measurements at the Relativistic
Heavy Ion Collider (RHIC) and the Large Hadron Collider (LHC), most notably
the observations of collective flow and jet quenching, have demonstrated that
the produced medium behaves as a strongly coupled
fluid~\cite{BRAHMS:2004adc,PHENIX:2004vcz,PHOBOS:2004zne,STAR:2005gfr,Muller:2012zq}.
A central objective of current research is to turn this qualitative picture
into a quantitative determination of the transport properties of the QGP.

Heavy quarks are particularly well suited for this
purpose~\cite{Rapp:2009my,Svetitsky:1987gq}. Their large masses restrict their
production to primordial hard scatterings, so that charm and bottom quarks
subsequently witness the entire space-time evolution of the
medium~\cite{Cao:2013ita}, and their interactions with the QGP constituents
remain imprinted on the spectra of the final heavy-flavor
hadrons~\cite{Rapp:2009my,He:2011qa}. Two classes of observables have proven
especially informative~\cite{ALICE:2012ab,ALICE:2018lyv,Yang:2023rgb}: the
nuclear modification factor $R_{AA}$, which encodes the medium-induced
spectral modification and the average in-medium energy loss, and the
anisotropic flow coefficients $v_n$, which carry sensitivity to the geometry
and collective dynamics of the expanding medium.

A quantitative description of the bulk medium is provided by relativistic
viscous hydrodynamics~\cite{Heinz:2013th,Gale:2013da,Florkowski:2017olj}, which,
when supplemented with fluctuating initial conditions on an event-by-event
basis, accounts for a broad range of soft-hadron
observables~\cite{Shen:2014vra,Song:2010mg}. Fluctuations in the positions of
the participating nucleons and in the deposited entropy generate
event-dependent spatial eccentricities that the hydrodynamic evolution
converts into final-state momentum
anisotropies~\cite{Alver:2010gr,Teaney:2010vd}. Within this picture, the
triangular flow $v_3$ originates predominantly from initial-state
fluctuations, while $v_4$ receives additional contributions from nonlinear
hydrodynamic response~\cite{ALICE:2011ab,Gardim:2011xv}.

The propagation of heavy quarks through the QGP is commonly described within
the Langevin framework, in which drag and stochastic forces encode the
medium-induced energy loss and momentum
diffusion~\cite{Svetitsky:1987gq,Moore:2004tg,Cao:2013ita,Chen:2017duy}.
Combined with coalescence and fragmentation hadronization, this approach has
been successfully applied to heavy-flavor observables ranging from $D$ mesons
to exotic multi-heavy states~\cite{Chen:2022X3872,Chen:2022Bc,Chen:2024DoublyCharmed,Li:2025OmegaCCC}.
Most of these applications, however, employ azimuthally averaged, smooth
medium backgrounds, and the influence of the event-by-event substructure of
the QGP on heavy-flavor observables remains comparatively
unexplored~\cite{Prado:2017ebe}. Indications of its importance already exist
in the quarkonium sector: the measured triangular flow of $J/\psi$ mesons has
been described by incorporating QGP fluctuations~\cite{Zhao:2022CharmoniumV3},
whereas bottomonium states respond only weakly, a consequence of their large
binding energies~\cite{Liu:2026cpg}. For open heavy-flavor, the
fluctuation-driven higher-order harmonics---in particular the fourth-order
coefficient $v_4$---remain an open quantitative question.

In the present work, we address this question for charm quarks in Pb-Pb
collisions at $\sqrt{s_{NN}} = 5.02$~TeV. We couple the iEBE-VISHNU
event-by-event viscous hydrodynamic framework~\cite{Shen:2014vra} to a
Langevin transport model~\cite{Cao:2013ita,Chen:2017duy} with instantaneous
coalescence hadronization, and compute the $D$ meson nuclear modification
factor $R_{AA}$ together with the anisotropic flow coefficients $v_2$, $v_3$,
and $v_4$. The results for $R_{AA}$, $v_2$, and $v_3$ are confronted with the
available experimental measurements, while the $v_4$ calculation constitutes a
prediction for future higher-order heavy-flavor flow measurements. The paper
is organized as follows. In Sec.II, we describe the
hydrodynamic medium, the Langevin transport model, and the hadronization
procedure. Section III presents our results for $R_{AA}$ and the
flow harmonics in comparison with data. We summarize and conclude in
Sec. IV.

\section{Theoretical Framework}
\label{sec:framework}

\subsection{Langevin Transport and Hadronization}
\label{sec:langevin}

The propagation of charm quarks through the fluctuating, rapidly expanding
medium is simulated within a Langevin
approach~\cite{Svetitsky:1987gq,Moore:2004tg,Cao:2013ita}, in which the
collisional interactions with the medium constituents are encoded in a
deterministic drag force and a stochastic noise
term~\cite{Moore:2004tg,Rapp:2009my}. The time evolution of the charm quark
momentum $\mathbf{p}$ is governed by the stochastic differential
equation~\cite{Cao:2013ita,Cao:2015hia,Chen:2017duy}
\begin{equation}
     \frac{d\mathbf{p}}{dt} = -\eta(p, T)\,\mathbf{p} + \boldsymbol{\xi}(t) +\boldsymbol{f}_g.
    \label{eq:langevin}
\end{equation}
Here, $\eta(p,T)$ denotes the drag coefficient, while $\boldsymbol{\xi}(t)$
represents the stochastic force arising from thermal scattering off the medium.
Assuming momentum-independent white noise, its two-point correlation function
reads~\cite{Svetitsky:1987gq,Kubo:1966fyg}
\begin{equation}
    \langle \xi_i(t)\,\xi_j(t') \rangle = \kappa(T)\,\delta_{ij}\,\delta(t-t'),
    \label{eq:noise}
\end{equation}
where $i,j$ label the Cartesian components and $\kappa(T)$ is the momentum
diffusion coefficient. We adopt the drag coefficient in the form
\begin{equation}
\eta(p,T)=\frac{\kappa(T)}{2\,T\,E(p)},
\qquad
E(p)=\sqrt{p^2+m_c^2},
\end{equation}
with $m_c=1.5~\mathrm{GeV}$, which respects the Einstein relation
$\kappa=2ET\eta$~\cite{Svetitsky:1987gq,Kubo:1966fyg,Moore:2004tg} and retains
a momentum dependence only through the relativistic energy $E(p)$. In addition to the collisional processes encoded in the drag and noise terms,
medium-induced gluon radiation is incorporated as a recoil force $\boldsymbol{f}_g$ in the
Langevin equation~\cite{Yang:2023rgb}, where the spectrum of the radiated gluons is given by the perturbative QCD calculations~\cite{Zhang:2003wk}. In the low- and moderate-$p_T$ region of the following calculations, the evolution of heavy quark momentum is dominated by the drag and noise terms.

The overall strength of the charm quark coupling to the QGP is controlled by
the spatial diffusion coefficient $D_s$, which is related to the momentum
diffusion coefficient via $2\pi T D_s = 4\pi T^3/\kappa$~\cite{Moore:2004tg,Cao:2013ita,Altenkort:2023oms}.
In the event-by-event hydrodynamic background, the transport coefficients are
evaluated at the local temperature $T(x,y,\tau)$ sampled along each charm quark
trajectory. In the deconfined region, $T>T_c=0.16~\mathrm{GeV}$, we vary the
diffusion coefficient within $1\leq 2\pi T D_s \leq 3$ to bracket the
theoretical uncertainty in its extraction; this band is compatible with
lattice-QCD computations and with phenomenological constraints inferred from
heavy-flavor observables~\cite{He:2011qa,Altenkort:2023oms,Xu:2017obm}. In the
temperature window $T_{\mathrm{freeze}}<T\leq T_c$, with
$T_{\mathrm{freeze}}=0.12~\mathrm{GeV}$, we adopt $2\pi T D_s=8$. 

The initial transverse-momentum distribution of charm quarks is taken from
FONLL calculations for $pp$
collisions~\cite{Cacciari:1998it,Cacciari:2012ny}, and their spatial distribution is proportional to the binary nucleon-nucleon
collision density $N_{\rm coll}(x,y)$ generated by a Monte Carlo Glauber
model~\cite{Miller:2007ri}. Cold-nuclear-matter effects modify the initial charm quark
distribution through a multiplicative shadowing factor $S_{\rm shadow}$,
evaluated as the thickness-weighted average of the EPS09 gluon shadowing with
the EPS09 package~\cite{Eskola:2009uj,Helenius:2012wd}. The factor is
centrality dependent, with $S_{\rm shadow}=0.72$ and $0.79$ for 0--10\% and
30--50\% central Pb-Pb collisions, respectively. Starting from these initial
conditions, the charm quarks are evolved according to
Eq.~(\ref{eq:langevin}) in the fluctuating hydrodynamic medium.

The conversion of charm quarks into open-charm mesons proceeds via an
instantaneous quark-coalescence
mechanism~\cite{Fries:2003vb,Greco:2003xt,Greco:2003vf}. The charm quark recombines with a nearby thermal light
antiquark from the surrounding fluid cell, yielding the $D$ meson momentum
distribution
\begin{equation}
\begin{split}
    \frac{dN_D}{d\mathbf{P}} = \int & \frac{d\mathbf{p}_1}{(2\pi)^2}\,\frac{d\mathbf{p}_2}{(2\pi)^2}\,
    f_c(\mathbf{p}_1)\,f_q(\mathbf{p}_2) \\
    & \times W_D(\mathbf{q})\,\delta^{(2)}(\mathbf{P}-\mathbf{p}_1-\mathbf{p}_2),
\end{split}
\label{eq:coalescence}
\end{equation}
where $\mathbf{P}$ is the $D$ meson momentum, $f_c$ the charm quark momentum
distribution at hadronization obtained from the Langevin evolution, and $f_q$
the light-quark momentum distribution, taken to be of Fermi-Dirac form. The
momentum-space Wigner function
$W_D(\mathbf{q}) = N_0 \exp(-\sigma^2\mathbf{q}^2)$ encodes the coalescence
probability in terms of the relative momentum $\mathbf{q}$ between the charm
and light quarks in the rest frame of the formed
meson~\cite{Xu:2025ivv}, with $N_0$ the normalization factor and the width
parameter $\sigma$ is determined via the wave function of $D$ meson~\cite{Chen:2022X3872}.

\subsection{Event-by-Event Hydrodynamics}

To simulate the spacetime evolution of the hot QCD medium produced in Pb-Pb collisions at $\sqrt{s_{NN}}=5.02$~TeV, we employ the iEBE-VISHNU event-by-event framework~\cite{Shen:2014vra}. The hydrodynamic evolution is performed within a longitudinally boost-invariant $(2+1)$-dimensional framework. The fluctuating initial conditions of the hydrodynamic model are constructed using the Monte Carlo Glauber model implemented in the \texttt{superMC} module~\cite{Broniowski:2007nz,Alver:2008aq,Loizides:2014vua}. The local entropy density profile $s(x,y)$ is parameterized as a linear combination of the wounded-nucleon and binary-collision distributions:
\begin{equation}
    s(x,y) = s_0 \left[(1-\alpha)\,\frac{N_{\mathrm{part}}(x,y)}{2} + \alpha\,N_{\mathrm{coll}}(x,y)\right],
\end{equation}
where $s_0$ is the overall normalization factor tuned to reproduce the experimentally measured charged-particle multiplicity within each centrality class, and $N_{\mathrm{part}}(x,y)$ and $N_{\mathrm{coll}}(x,y)$ denote the local density profiles of participant nucleons and binary nucleon-nucleon collisions, respectively~\cite{Miller:2007ri}. The mixing parameter is tuned to reproduce the measured multiplicity with $\alpha = 0.118$ at LHC energies, which dynamically regulates the relative contribution between the two density components to replicate the experimental central-to-peripheral multiplicity scaling~\cite{ALICE:2015juo}. 

The subsequent macroscopic space-time evolution of the strongly coupled bulk medium is governed by $(2+1)$-dimensional relativistic viscous hydrodynamics, described by the energy-momentum conservation laws~\cite{Song:2007ux}:
\begin{equation}
    \partial_{\mu}T^{\mu\nu}=0.
\end{equation}
In this study, we implement a constant specific shear viscosity $\eta/s = 0.08$ and neglect the bulk viscous pressure. The hydrodynamic expansion starts at a proper time $\tau_0 = 0.6~\mathrm{fm}/c$ and utilizes the s95p-PCE equation of state~\cite{Huovinen:2009yb}. As the system expands and cools to a decoupling energy density of $E_{\mathrm{dec}} = 0.18~\mathrm{GeV/fm}^3$, the fluid is converted into microscopic hadrons via the Cooper-Frye freeze-out formalism~\cite{Cooper:1974mv}. In this work, we generate { 1000} fluctuating events for each centrality class, which will be used for heavy quark evolutions.

\subsection{Heavy-Flavor Observables}
The anisotropic emission pattern of final-state particles in the fluctuating
medium is quantified by the Fourier coefficients $v_n$ of the azimuthal
distribution~\cite{Ollitrault:1992bk,Voloshin:1994mz},
\begin{equation}
    \frac{dN}{d\phi}
    \propto
    1+2\sum_{n=1}^{\infty}
    v_n\cos\!\bigl[n(\phi-\Psi_n)\bigr],
    \label{eq:vn}
\end{equation}
where $\phi$ is the azimuthal angle of the particle and $\Psi_n$ the
event-plane angle of the $n$-th harmonic. In each event $e$, the event planes
$\Psi_{n,e}^{\rm soft}$ are reconstructed separately for every harmonic from
the azimuthal distributions of the final-state soft
hadrons~\cite{Poskanzer:1998yz}. The elliptic flow is driven
primarily by the elliptic geometry of the initial overlap region, whereas the
triangular flow arises predominantly from event-by-event fluctuations of that
geometry. The fourth harmonic, in turn, receives both a linear response to
the initial fourth-order eccentricity $\epsilon_4$ and a nonlinear
contribution coupled to the elliptic flow~\cite{Teaney:2012ke}.

The differential flow coefficients of $D$ mesons in a given
transverse-momentum interval are evaluated with respect to the soft-hadron
event plane of the same event,
\begin{equation}
v_n^D(p_T)
=
\frac{
\displaystyle
\sum_e\sum_{i\in e}
\cos\left[
n\left(
\phi_{D,i}-\Psi_{n,e}^{\mathrm{soft}}
\right)
\right]
}{
\displaystyle
\sum_e N_{D,e}(p_T)
},
\label{eq:vnD}
\end{equation}
with $n=2,3,4$. Here $i$ labels the $D$ mesons in event $e$, $\phi_{D,i}$
denotes the azimuthal angle of the $i$-th $D$ meson, and $N_{D,e}(p_T)$ is
the number of $D$ mesons of that event within the selected $p_T$ interval;
the sums extend over all $D$ mesons and all events entering the analysis.
Together with the nuclear modification factor $R_{AA}$, these coefficients
quantify the momentum-space anisotropy and the medium-induced modification of
$D$ meson production in Pb-Pb collisions.

\section{Results in Heavy Ion Collisions}

\subsection{Light-hadron observables}
\label{sec:light}

\begin{figure}[htbp]
\centering
\includegraphics[width=0.8\linewidth]{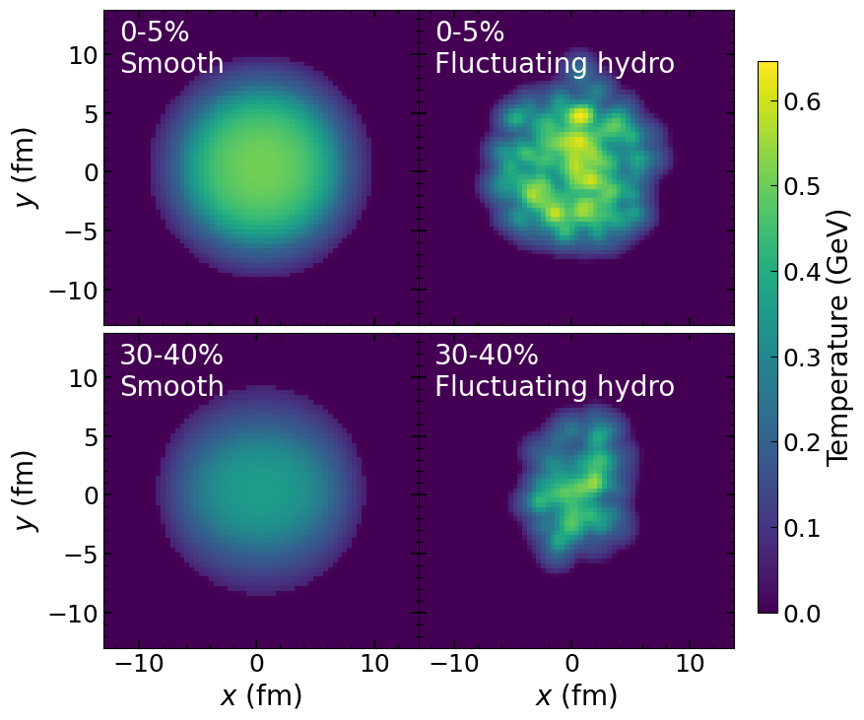}
\caption{Initial transverse temperature profiles at $\tau_0=0.6~\mathrm{fm}/c$
for the smooth (left) and a representative event-by-event fluctuating (right)
hydrodynamic background in 0--5\% (top) and 30--40\% (bottom) Pb-Pb
collisions at $\sqrt{s_{NN}}=5.02$~TeV.}
\label{fig:hydro_maps}
\end{figure}

Before turning to the heavy-flavor results, we validate the hydrodynamic
medium that drives the charm-quark transport against soft-hadron
measurements, since the light-hadron observables are obtained directly from
the hydrodynamic evolution.
Figure~\ref{fig:hydro_maps} shows the initial transverse temperature profiles
of the smooth and a representative fluctuating background in the 0--5\% and
30--40\% centrality classes. The fluctuating medium exhibits localized hot
spots that originate from event-by-event fluctuations of the initial entropy
deposition. The overall normalization of the entropy deposition is fixed by
the final charged-particle multiplicity: Table~\ref{tab1} compares the mean
charged-particle pseudorapidity densities $\langle dN_{\rm ch}/d\eta\rangle$,
averaged over 1000 fluctuating events, with the corresponding ALICE
measurements~\cite{ALICE:2019hno}. The calculations reproduce the
experimental multiplicities within uncertainties over the full centrality
range considered.

\begin{table}[htbp]
    \centering
    \caption{Mean charged-particle pseudorapidity density at midrapidity
    ($\vert\eta\vert<0.5$) from the hydrodynamic calculations compared with
    the ALICE measurements~\cite{ALICE:2019hno} in Pb-Pb collisions at
    $\sqrt{s_{NN}} = 5.02$~TeV.}
    \label{tab1}
\begin{tabular}{ccc}
\hline
Centrality &$\langle dN_{ch}/d\eta\rangle_{\rm th}$&$\langle dN_{ch}/d\eta\rangle_{\rm exp}$\\
\hline
0-5\%& 1996.401& $1943\pm 56$ \\
5-10\%& 1622.881& $1587\pm 47$ \\
30-40\%& 505.613& $512\pm15$\\
50-60\%& 172.0854& $183\pm8$\\
80-90\%& 16.527& $17.5\pm1.8$\\
\hline
\end{tabular}
\end{table}

\begin{figure}[htbp]
\centering
\includegraphics[width=0.8\linewidth]{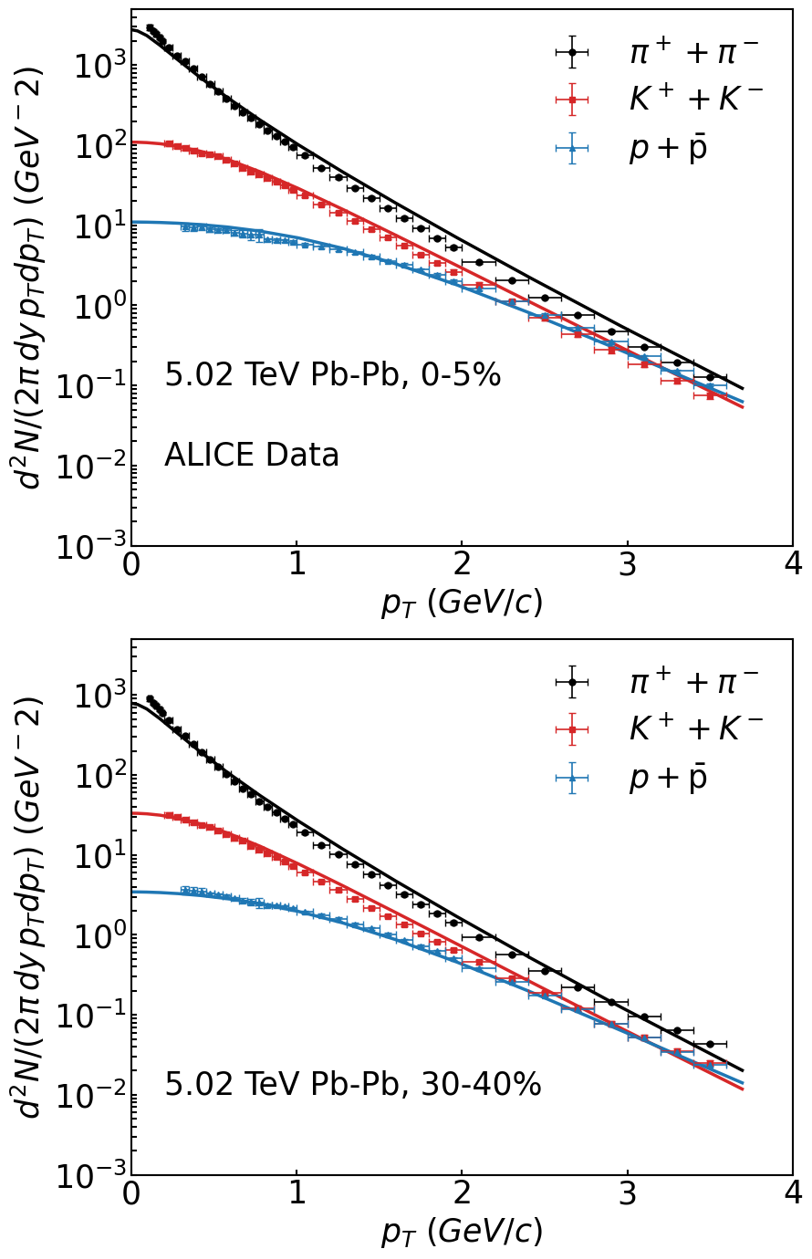}
\caption{Transverse-momentum spectra of charge-averaged identified hadrons in
0--5\% (top) and 30--40\% (bottom) Pb-Pb collisions at
$\sqrt{s_{NN}}=5.02$~TeV. Solid curves denote the hydrodynamic calculations
and symbols the ALICE measurements~\cite{ALICE:2019hno}.}
\label{lab-figdN}
\end{figure}

Figure~\ref{lab-figdN} presents the transverse-momentum spectra of the
charge-averaged identified hadrons, $(\pi^++\pi^-)/2$, $(K^++K^-)/2$, and
$(p+\bar p)/2$, in the 0--5\% and 30--40\% centrality classes, computed from
the ensemble of fluctuating events and compared with the experimental data.
The calculations reproduce the spectral shapes and the mass ordering of the
identified hadrons in the soft $p_T$ regime; the remaining deviations are
consistent with the absence of a late-stage hadronic cascade (e.g., UrQMD),
whose rescatterings further reshape the spectra.

\begin{figure*}[htp]
\centering
\subfloat[0-5\%]{
\includegraphics[width=0.9\linewidth]{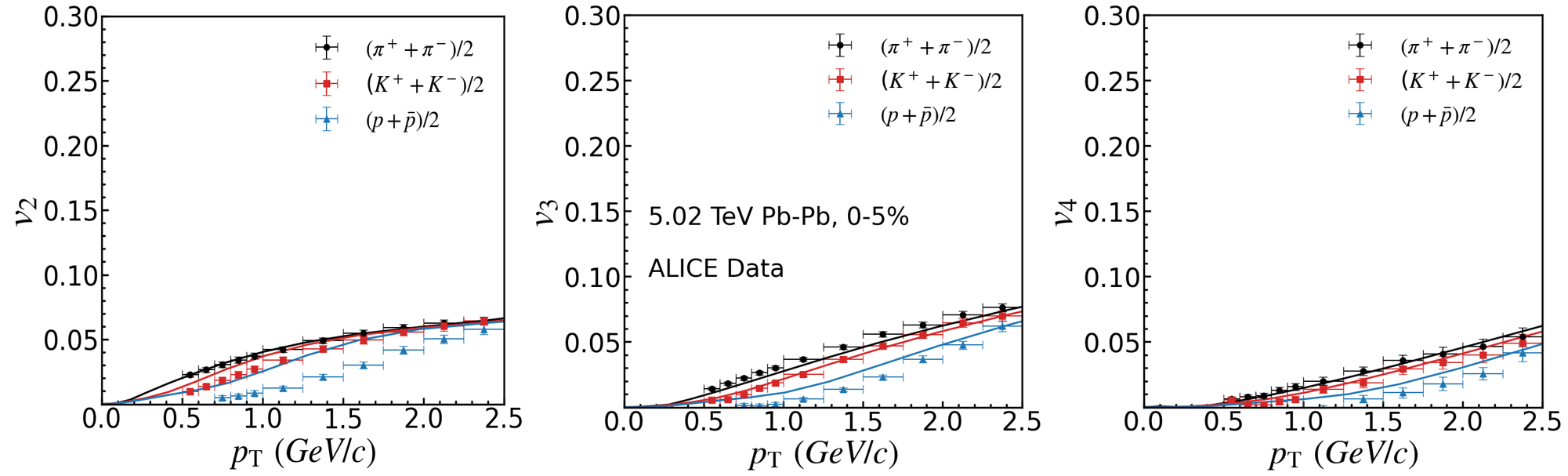}
}
\hfill
\subfloat[30-40\%]{
\includegraphics[width=0.9\linewidth]{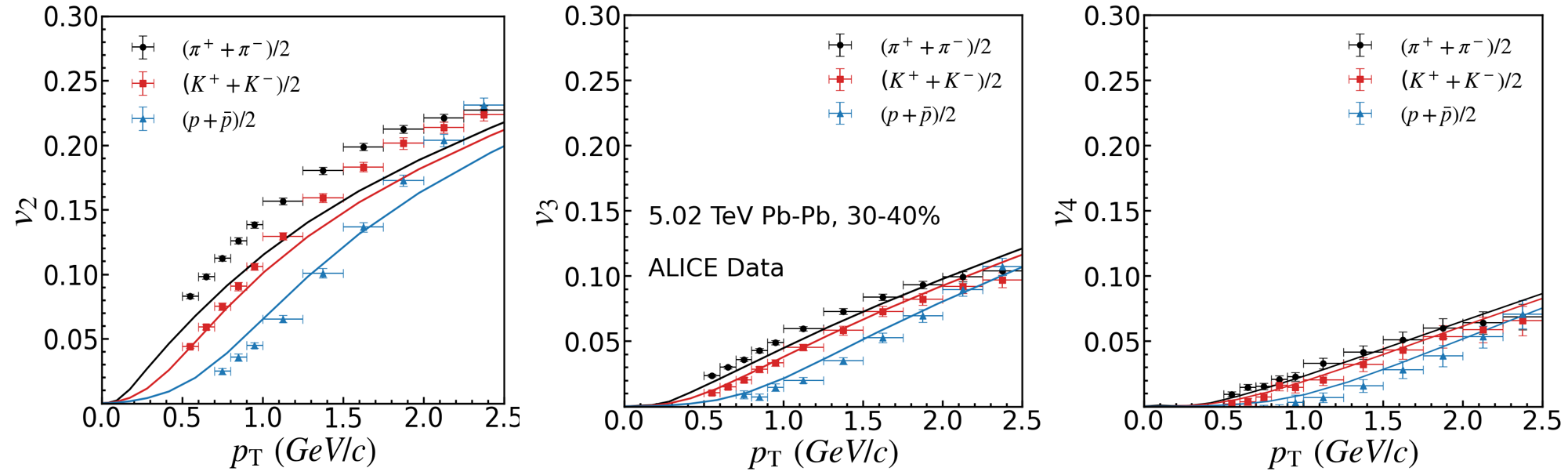}
}
\caption{Anisotropic flow coefficients $v_2$, $v_3$, and $v_4$ of
charge-averaged identified hadrons in 0--5\% (top) and 30--40\% (bottom)
Pb-Pb collisions at $\sqrt{s_{NN}}=5.02$~TeV, compared with ALICE
measurements~\cite{ALICE:2018yph}.}
\label{lab-figflow}
\end{figure*}

The anisotropic flow coefficients $v_2$, $v_3$, and $v_4$ of the identified
light hadrons are shown in Fig.~\ref{lab-figflow} for the same two centrality
classes. The calculations reproduce the measured $p_T$ dependence and the
hierarchy $v_2 > v_3 > v_4$ of the data. Deviations remain visible, most
noticeably for $v_2$ in the 30--40\% centrality class, and can again be
ascribed in part to the omitted hadronic afterburner, since hadronic
rescattering modifies both the momentum distributions and the anisotropic
flow of identified hadrons. Overall, this level of agreement demonstrates
that the fluctuating iEBE-VISHNU backgrounds capture the bulk collective
dynamics required for the heavy-flavor transport study presented below.

\subsection{Heavy-flavor results}
\label{sec:hf}

\begin{figure}[htbp]
\centering
\includegraphics[width=0.8\linewidth]{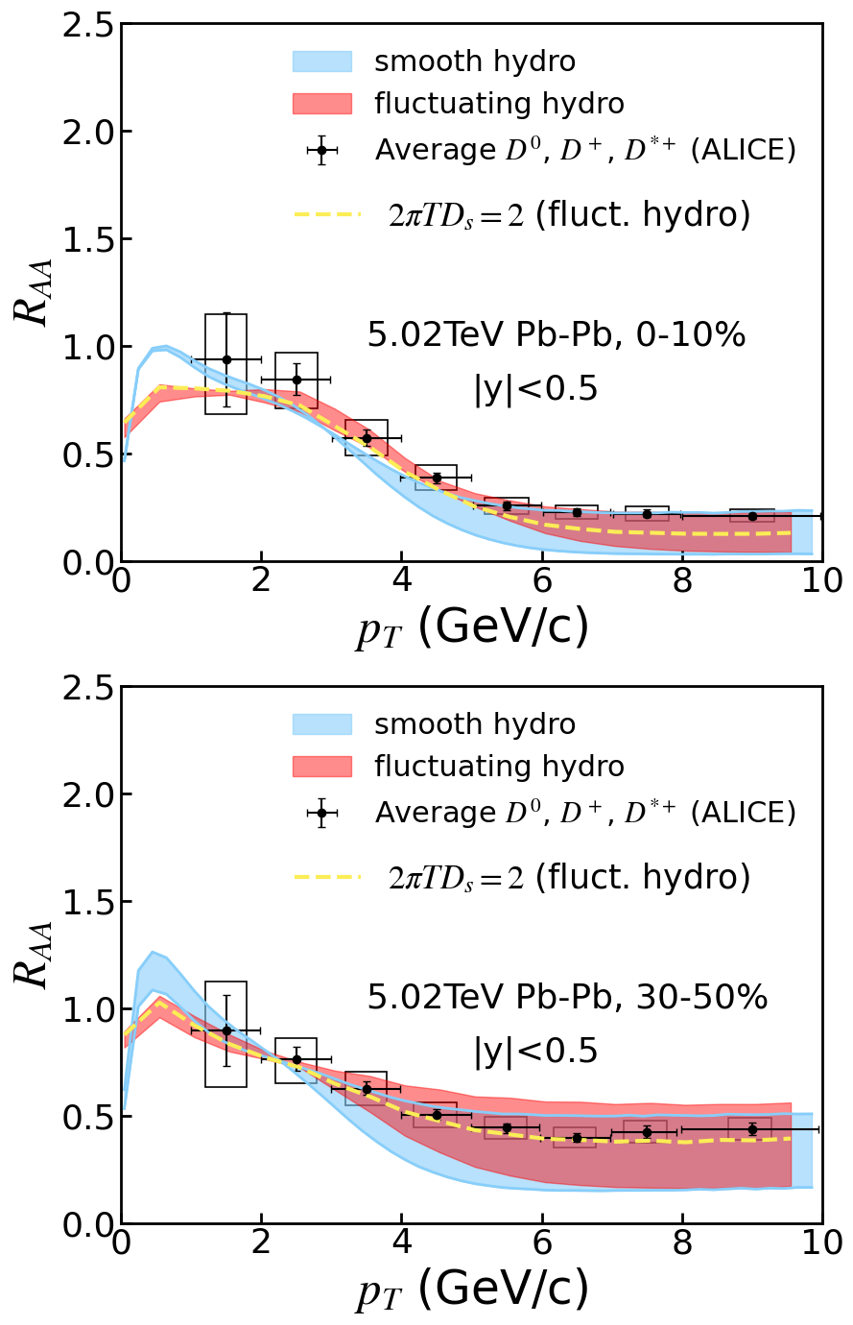}
\caption{Nuclear modification factors of $D$ mesons in Pb-Pb collisions at $\sqrt{s_{NN}}=5.02$~TeV for 0-10\% (top) and 30-50\% (bottom) centrality classes. The shaded bands represent the calculations with the spatial diffusion coefficient varied within $1\leq2\pi T D_s\leq3$ and the dashed line denotes the results of the fluctuating hydro for $2\pi T D_s = 2$. The results obtained in a smooth hydrodynamic background are also shown for comparison. Experimental data are taken from the ALICE Collaboration~\cite{ALICE:2021rxa}.}
\label{fig:RAA}
\end{figure}

Figure~\ref{fig:RAA} shows the $D$-meson nuclear modification factor $R_{AA}$
in the 0--10\% and 30--50\% centrality classes, compared with the ALICE
measurements~\cite{ALICE:2021rxa}. The bands of the $R_{AA}$ corresponds to two values of the spatial diffusion coefficients, $2\pi T D_s=1$ and $3$, which characterize the coupling strength between heavy quarks and the bulk medium. The dashed line is the calculation with the middle value $2\pi T D_s=2$. Smooth and fluctuating medium are considered respectively. 
The pronounced suppression at intermediate
and high $p_T$ reflects the medium-induced modification of the charm-quark
momentum distributions accumulated over the in-medium evolution. The
event-by-event and smooth-background calculations follow similar trends,
with residual differences concentrated at low $p_T$; initial-state
fluctuations thus exert only a moderate, momentum-dependent influence on the
azimuthally averaged $R_{AA}$. The agreement with the data demonstrates that
the coupled hydrodynamic--Langevin framework provides a realistic description
of charm-quark energy loss and hadronization at
$\sqrt{s_{NN}} = 5.02$~TeV.

\begin{figure*}[htbp]
\centering
\includegraphics[width=0.9\linewidth]{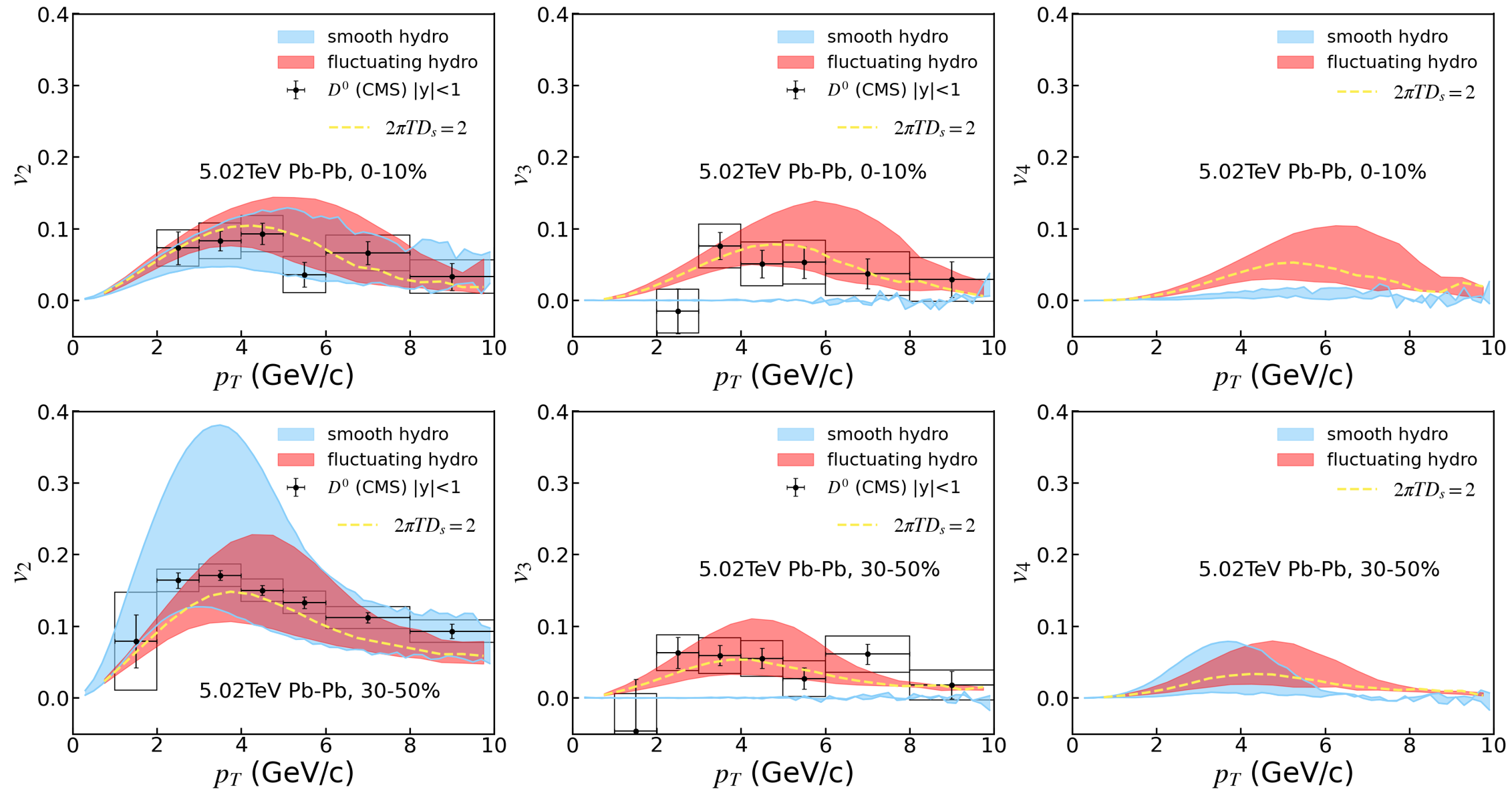}
\caption{Anisotropic flow coefficients $v_2$, $v_3$, and $v_4$ of $D$ mesons in 0-10\% (top) and 30-50\% (bottom) Pb-Pb collisions at $\sqrt{s_{NN}}=5.02$~TeV. The shaded bands represent the results obtained with $1 \leq 2\pi T D_s \leq 3$, while the curves correspond to $2\pi T D_s = 2$. Results obtained using a smooth (non-fluctuating) hydrodynamic background are also shown for comparison. Available CMS measurements of prompt $D^0$-meson $v_2$ and $v_3$ are included~\cite{CMS:2017vhp}.}
\label{fig:flow}
\end{figure*}

The anisotropic flow coefficients $v_2$, $v_3$, and $v_4$ of $D$ mesons are
presented in Fig.~\ref{fig:flow} for the same two centrality classes,
together with the CMS measurements of prompt $D^0$-meson $v_2$ and
$v_3$~\cite{CMS:2017vhp} and our predictions for $v_4$. To separate the
contributions of the average collision geometry from those of initial-state
fluctuations, the smooth-background results are displayed alongside the
event-by-event bands.

The elliptic flow is compatible with the data within uncertainties over most
of the $p_T$ range in both centrality classes. In the 30--50\% class, the
event-by-event $v_2$ falls visibly below the smooth-background result. Since
the $D$-meson $v_2$ is built up through both the hydrodynamic response to the
initial geometry and the subsequent charm-quark transport, this reduction
indicates that fluctuations affect both stages. A similar suppression was
reported in Ref.~\cite{Singh:2023smw}, where it was traced to the
competition between two fluctuation-driven effects: enhanced local pressure
gradients, which strengthen the flow response, and an increased decorrelation
between the heavy- and light-flavor event planes, which reduces the $v_2$
measured relative to the soft event plane. The dominance of the latter effect
found in that study offers a natural interpretation of the lower
event-by-event $v_2$ in Fig.~\ref{fig:flow}.

The triangular flow $v_3$ exhibits the opposite pattern. The smooth-background
calculation yields an essentially vanishing $v_3$, as the azimuthally
averaged geometry carries no triangular eccentricity. In the fluctuating
events, finite initial triangular eccentricities drive a triangular response
of the expanding medium~\cite{Katz2020,2xt8-bh6x}, which is transmitted to
the charm quarks through their interactions with the QGP and survives
hadronization as a finite $D$-meson $v_3$. The result is broadly consistent
with the CMS measurements within uncertainties, confirming that initial-state
fluctuations constitute the primary source of $D$-meson triangular flow.

Turning to the fourth harmonic $v_4$, the smooth-background calculation yields only
a small $v_4$ in the 0--10\% centrality class, whereas the event-by-event
calculation predicts a more pronounced signal, particularly at intermediate
and high $p_T$. In the 30--50\% class, both calculations produce sizable
values but with distinct shapes: the smooth-background result peaks at lower
$p_T$, while the event-by-event result develops a broader distribution
extending to higher $p_T$. Event-by-event fluctuations thus modify the
fourth-order flow of $D$ mesons in a centrality- and $p_T$-dependent manner.

These features follow naturally from how the fourth harmonic is generated.
In the bulk medium, $v_4$ receives a linear response to the initial
eccentricity $\epsilon_4$ and a nonlinear contribution from the coupling of
the elliptic flow to itself during the hydrodynamic
evolution~\cite{Teaney:2012ke,Yan:2015jma}. The $D$-meson $v_4$ inherits both
sources through the charm-quark coupling to the corresponding medium
anisotropies. A nonlinear correlation between the $v_2$ and $v_4$ of $D$
mesons has indeed been observed in event-shape engineering
analyses~\cite{Sambataro:2022sns}. The enhanced and broadened $v_4$ in the
event-by-event calculation can accordingly be attributed to the additional
fluctuation-driven anisotropies transferred to the charm quarks during the
hydrodynamic evolution; in central collisions, where the average geometric
anisotropies are small, fluctuations provide the dominant source of the
finite $D$-meson $v_4$. The relative weights of the average geometry and the
fluctuations in the $D$-meson $v_4$ therefore depend on both collision
centrality and transverse momentum.

\section{Summary}
\label{sec:summary}

In summary, we have studied charm-quark transport in Pb-Pb collisions at
$\sqrt{s_{NN}} = 5.02$~TeV by embedding a Langevin transport model with
instantaneous coalescence hadronization into the event-by-event viscous
hydrodynamic backgrounds provided by iEBE-VISHNU. The medium evolution was
validated against soft-hadron multiplicities, transverse-momentum spectra,
and anisotropic flow harmonics, ensuring a realistic fluctuating background
for the heavy-flavor stage of the calculation.

Our results establish a clear separation of roles among the heavy-flavor
observables. The azimuthally averaged nuclear modification factor $R_{AA}$
is governed primarily by the average in-medium energy loss and
cold-nuclear-matter effects, and acquires only moderate, momentum-dependent
modifications from event-by-event fluctuations. The anisotropic flow
coefficients, in contrast, respond sensitively to the fluctuating initial
geometry. The calculations reproduce the measured $D$-meson $v_2$, and the
triangular flow $v_3$---absent in a smooth background---is generated entirely
by initial-state fluctuations, in agreement with the available CMS
measurements. This demonstrates that charm quarks acquire the collectivity
imprinted by the event-by-event substructure of the QGP.

Finally, our predictions for the fourth-order flow coefficient $v_4$ extend
this picture to higher harmonics, where the interplay between the average
geometry and initial-state fluctuations varies with centrality and transverse
momentum. A future measurement of the $D$-meson $v_4$ would test these
predictions and provide an additional constraint on the coupling of heavy
quarks to the fluctuating medium.

\vspace{1cm}
{\bf Acknowledgements:} 
This work is supported
by the National Natural Science Foundation of China
(NSFC) under Grant Nos. 12575149 and 12175165.

\bibliographystyle{apsrev4-2}
\bibliography{ref}
\end{document}